# Velocity Discretization for Lattice Boltzmann Method for Noncontinuum Bounded Gas Flows at the Micro and Nanoscale

Yong Shi [*]

Department of Mechanical, Materials and Manufacturing Engineering, University of Nottingham Ningbo China, Ningbo 315100, China

**ABSTRACT**

The lattice Boltzmann (LB) method intrinsically links to the Boltzmann equation with the Bhatnagar-Gross-Krook collision operator; however, it has been questioned to be able to simulate noncontinuum bounded gas flows at the micro and nanoscale, where gas moves at a low speed but has a large Knudsen number. In this article, this point has been verified by simulating Couette flows at moderate and large Knudsen numbers (e.g., $Kn=10$ and $Kn=100$) by the linearized LB models based on the popular half-space Gaussian Hermite (HGH) quadrature. The underlying cause for poor accuracy of these conventional models is analyzed in light of numerical evaluation of the involved Abramowitz functions. A different thought on velocity discretization is then proposed using Gauss Legendre (LG) quadrature. Strikingly, the resulting GL-based LB models have achieved high accuracy in simulating Couette flows in the strong transition and even free molecular flow regimes. The numerical study in this article reveals an essentially different while workable way in constructing LB models for simulating micro and nanoscale low-speed gas flows with strong noncontinuum effects.

---

[*] Tel: +86 (0)574 88186413; Fax: +86 (0)574 88180715
E-mail addresses: Yong.Shi@nottingham.edu.cn.

## I. INTRODUCTION

The lattice Boltzmann (LB) method emerges as a modification of lattice gas cellular automata to remove statistical noise caused by Boolean operation [1,2]. Its classical version mimics particle dynamics as a streaming-collision process based on the distribution function on a regular lattice at a finite number of discrete velocities [2–4]. Although greatly simplified, such particle dynamics well produced a large variety of macroscopic flow phenomena, and the LB method has achieved tremendous successes in simulating continuum flows over the last thirty years [1–4]. Historically, one significant progress in the development of the LB method is introduction of the Bhatnagar-Gross-Krook (BGK) assumption in its framework for particle collision, which led to the LB BGK models [5,6]. These models were later reconstructed as special numerical schemes of the Boltzmann BGK equation [7–10]. Such an intrinsic link provides a direct interpretation of the LB method in the context of the kinetic theory of gases, and aroused broad interests in applying the LB method to simulate gas flows at different Knudsen numbers, $Kn$ [10–25].

As a numerical approach for simulating gas flows, discretization of the velocity space is of crucial importance for the accuracy of the LB method. The early work [7,8] demonstrates appropriate discretization of the velocity space can be analytically derived using the full-space Gauss Hermite (FGH) quadrature. The demonstration is based on two premises: 1. The equilibrium distribution in the LB method is expanded as a Taylor series in the velocity space, under the small Mach number condition; 2. The LB equation and distribution function are expanded by the Chapman-Enskog series. In so doing, the moments of interest are reduced to integrals of polynomials in the velocity space with respect to a weight function, $\exp\left[-\mathbf{c}^2/(2RT)\right]$, where $\mathbf{c}$, $R$ and $T$ are the particle velocity, gas constant and temperature. They can be exactly evaluated order by order using the FGH quadrature at increasing algebraic precision (AP) [26]. However, success in the LB modeling along this avenue is limited to continuum flows as the Chapman-Enskog expansion lacks general validity for noncontinuum flows beyond the Navier-Stokes order [27].

Another thought on discretization of the velocity space in the LB method is based on Grad's theory [28], where the distribution function is expanded in terms of the Hermite polynomials in the velocity space [9,10]. Thanks to the mutual orthogonality of these polynomial terms, such an expanded distribution function possesses an important property–its leading moments, e.g., up to the $N^{th}$ order, will remain unchanged when the higher-order terms $(\geq N+1)$ in the expansion are truncated. Therefore, for a flow specified by the $N$ moments, its distribution function can be further simplified as a finite series only including the first $N$ Hermite polynomials, i.e., $f^N$ [10]. Again, the moments based on $f^N$ are the integrals of polynomials in the velcity space and evaluated by the FGH quadrature. In Refs [9,10], a hierarchy of the LB models was constructed in line with this rationale, and solved the truncated distribution functions in terms of the FGH-based discrete velocities. The results contain flow details equivalent to those of the macroscopic equations based on the $N$ moments. In this sense, the LB method is applicable to simulate noncontinuum flows; its accuracy is the same as that of the corresponding moment equations [10].

In the LB framework based on Grad's expansion, its accuracy is controlled by AP of the GH quadrature. Nonetheless, when those FGH-based LB models were applied to simulate noncontinuum flows bounded by solid surfaces, it was found that their accuracy is not monotonically improved with the increasing AP (corresponding to the increasing quadrature nodes). On the contrary, variations of the LB accuracy oscillates between the models at even quadrature nodes and those at odd quadrature nodes [14-16]. This numerical phenomenon is attributed to the discontinuity of the distribution function in the velocity space caused by Maxwell diffusive boundary condition [16], and leads to the half-space Gauss Hermite (HGH) quadrature [29] being adopted in the recent LB literature [14-16, 17, 19-20, 24]. To be specific, the HGH quadrature in these studies is only employed to discretize the positive half of the velocity space, if it is one dimensional (1D), while the discrete velocities in the negative half are specified by origin symmetry. For multidimensional cases, tensor product is used to produce 2D or 3D discrete velocities based on the 1D results. In comparison to their FGH-based counterparts, the HGH-based LB models obtain more accurate results for bounded gas flows in the slip ($0.01 < Kn < 0.1$) and weakly transition ($0.1 < Kn < 1$) flow regimes [17,19-20].

Note that simulating bounded gas flows in the strong transition $(1 \leq Kn \leq 10)$ and free molecular $(Kn > 10)$ flow regimes is still a challenge in today's LB study. In these problems, the Knudsen layer occupies a large portion, if not all, of the entire flow domain, in which the distribution function behaves away from a polynomial in $\mathbf{c}$ [30]. Large numerical errors are therefore observed in the LB results, including those by the HGH-based LB models (e.g., in Ref. [19]). In recent years, these failures have cast wide doubt on the kinetic nature of the LB method and its applicability to simulate noncontinuum gas flow in micro and nanostructures. In this article, I tackle this issue focusing on velocity discretization for the LB modelling at large Knudsen numbers. The conventional HGH quadrature schemes at different AP are first used to construct a hierarchy of the linearized LB models. Couette flow at micro/nanoscale (i.e., a linear flow problem at a large Knudsen number) is simulated, and the accuracy of these HGH-based LB models is assessed in comparison to high-accuracy solutions to the linearized Boltzmann BGK equation [31] numerically solved by the method in Ref. [32]. It is found that the low accuracy of the LB results roots from the misuse of the HGH quadrature for evaluating the Abramowitz functions involved in the flows. As an alternative, a new way for velocity discretization is proposed based on the Gauss Legendre (GL) quadrature, in combination with the range truncation of the particle velocity and Möbius transformation. Dramatic improvements are made in the resulting LB simulations of Couette flows at $Kn=10$ and even at $Kn=100$.

The rest of this article is organized as follows: In Sec. II, the linearized Boltzmann BGK equation is introduced and applied to describe Couette flow. Based on this kinetic equation, a hierarchy of the HGH-based LB models at different AP is constructed and applied to simulate Couette flows at $Kn=10$ and $Kn=100$ in Sec III. In Sec IV, I discuss the numerical deficiencies of these HGH-based LB models for simulating gas flows at large Knudsen numbers, and propose the velocity discretization based on the GL quadrature. The latter gives rise to the GL-based LB models in Sec V, together with their numerical results for strongly noncontinuum Couette flows. Finally, I draw the conclusion in Sec. VI.

## II. LINEARIZED BOLTZMANN BGK EQUATION FOR COUETTE FLOW

**A. The linearized Boltzmann BGK equation**

In this article, all LB models are constructed based on the linearized Boltzmann BGK equation. This is mainly because Couette flow is a linear problem, without including the nonlinear advection. Its high-accuracy solutions to the linearized Boltzmann BGK equation are available in a wide range of the Knudsen number [31,32]. Moreover, the linearized Boltzmann BGK equation is also a good kinetic model for describing low-speed noncontinuum gas flows bounded in micro and nanostructures [32-37], which are usually subject to a large Knudsen number, but their Reynolds number and Mach number are negligibly small. As shown in Ref. [30], the linearized Boltzmann BGK equation is

$$\frac{\partial h}{\partial t} + \mathbf{c} \cdot \frac{\partial h}{\partial \mathbf{x}} = -\frac{1}{\tau}\left(h - h^{eq}\right), \tag{1}$$

where $t$, $\mathbf{x}$, $\mathbf{c}$ and $\tau$ represent time, particle position, velocity and relaxation time, respectively. $h$ and $h^{eq}$ are the distribution function in the linearized Boltzmann theory. They are actually the perturbations to the original distribution function $f$ and its local equilibrium $f^{eq}$ from the global

Maxwellian, $f_0$, of a quiescent gas (the velocity $\mathbf{u}_0 = 0$) at a constant density $\rho_0$ and temperature $T_0$, i.e.,

$$h = \frac{f}{f_0} - 1 \quad \text{and} \quad h^{eq} = \frac{f^{eq}}{f_0} - 1, \tag{2}$$

where

$$f_0 = \frac{\rho_0}{(2\pi RT_0)^{D/2}} \exp\left(-\frac{\mathbf{c}^2}{2RT_0}\right), \tag{3}$$

and $D$ is the dimensionality of the physical space. Under the isothermal conditions, $h^{eq}$ in Eq. (1) is formulated by

$$h^{eq} = \frac{\delta\rho}{\rho_0} + \frac{\mathbf{c}\cdot\mathbf{u}}{RT_0}. \tag{4}$$

$\delta\rho$ and $\mathbf{u}$ are the perturbations to the gas density and velocity, which can be computed as the moments of $h$,

$$\delta\rho = \int f_0 h \, d\mathbf{c}, \qquad \rho_0 \mathbf{u} = \int f_0 h \mathbf{c} \, d\mathbf{c}. \tag{5}$$

Note that although linear, Eq. (1) is an integro-differential equation in a high dimensional space spanned by $t$, $\mathbf{x}$ and $\mathbf{c}$. It is difficult to obtain its analytical solutions for general gas flow problems.

**B. Couette flow between two flat plates**

Couette flow is one of a few problems, for which the analytical solutions or high-accuracy numerical solutions to the linearized Boltzmann BGK equation are available at different Knudsen numbers [31,32]. It is thus chosen in this article as a test problem to examine the accuracy of the LB method for simulating noncontinuum flows bounded by flat solid walls.

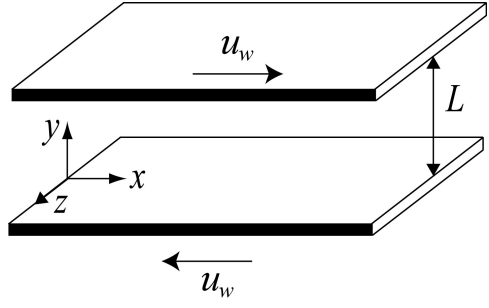

FIG. 1. Schematic of Couette flow. The origin of the coordinate system is located at the center of the bottom plate. It is fixed in the inertial frame.

Figure 1 illustrates Couette flow confined between two parallel flat plates, which are moving along their planes in the opposite directions but at the same speed, $u_w$. Under the isothermal conditions, this flow is described by Eqs. (1) and (4), subject to the Maxwell diffusive boundary conditions on the plate walls

$$\text{Bottom plate } (y=0): \quad h_w^0 = -\frac{c_x \cdot u_w}{c_s^2}, \quad \text{for } c_y > 0, \tag{6}$$

$$\text{Top plate } (y=L): \quad h_w^L = \frac{c_x \cdot u_w}{c_s^2}, \quad \text{for } c_y < 0, \tag{7}$$

where $c_x$ and $c_y$ are the components of $\mathbf{c}$ in the $x$ and $y$ directions, and the sound speed $c_s = \sqrt{RT_0}$. With the help of Eqs. (6) and (7), Eq. (1) is integrated along the characteristics line ($\mathbf{c} = d\mathbf{x}/dt$), leading to $h$ as

$$h = \begin{cases} h_w^0 \exp\left(-\frac{aY}{Kn}\frac{1}{\xi_y}\right) + \frac{a}{Kn}\frac{1}{\xi_y}\int_0^Y \exp\left[\frac{a(Y'-Y)}{Kn}\frac{1}{\xi_y}\right] \cdot h^{eq}(Y')\,dY', & \text{for } \xi_y > 0, \\ h_w^L \exp\left[\frac{a(1-Y)}{Kn}\frac{1}{\xi_y}\right] - \frac{a}{Kn}\frac{1}{\xi_y}\int_Y^1 \exp\left[\frac{a(Y'-Y)}{Kn}\frac{1}{\xi_y}\right] \cdot h^{eq}(Y')\,dY', & \text{for } \xi_y < 0, \end{cases} \tag{8}$$

where $a = \sqrt{\pi}/2$ and the Knudsen number $Kn = \sqrt{\pi RT_0}\,\tau/(\sqrt{2}L)$. The dimensionless velocities and coordinate are defined as $(\xi_x, \xi_y) = (c_x, c_y)/\sqrt{2RT_0}$ and $Y = y/L$, respectively. Substituting Eq. (8)

into Eq. (5) yields the dimensionless streamwise velocity (i.e., the component in the $x$ direction $U = u/u_w$),

$$U = \frac{1}{\sqrt{\pi}} \left\{ \mathscr{I}_0\left[\frac{a(1-Y)}{Kn}\right] - \mathscr{I}_0\left(\frac{aY}{Kn}\right) \right\} + \frac{1}{2Kn} \int_0^1 \mathscr{I}_{-1}\left(\frac{a|Y'-Y|}{Kn}\right) \cdot U(Y')\, dY'. \qquad (9)$$

In Eq. (9), the Abramowitz function $\mathscr{I}_n(x) = \int_0^{+\infty} \xi_y^n \exp\left(-\xi_y^2 - x/\xi_y\right) d\xi_y$ [38]. It is an integral over the positive half space of $\xi_y$. For convenience, I will denote its integrand as $I_n(\xi_y, x)$ in the following discussion. Equation (9) actually is the Fredholm integral equation of the 2$^{nd}$ kind, and can be numerically solved using the method in Ref. [32]. In Sections III and V, the high-accuracy streamwise velocities are given by solving Eq. (9) for Couette flows at $Kn = 10$ and $Kn = 100$ [31]. These solutions will be used as the benchmarks to assess the LB simulations based on the HGH and GL quadrature.

## III. LINEARIZED OFF-LATTICE LATTICE BOLTZMANN MODEL

The linearized LB models are constructed from Eq. (1) by discretizing its time, physical space and velocity space. To accommodate large discrete velocity spaces resulted from the HGH quadrature schemes at high AP, the linearized LB models in this article are formulated in the off-lattice framework [39].

### A. The lattice Boltzmann equation

The evolution equation of a linearized off-lattice LB model is obtained by integrating Eq. (1) over a time step, $\Delta t = t_{n+1} - t_n$. The trapezoid rule is applied to the integration of the collision term, while the advection term on the left-hand side of Eq. (1) is evaluated at $t_n$. All the involved spatial gradients are approximated by the finite difference schemes [39]. After several mathematical manipulations, the evolution equation is obtained

$$H_i^{n+1} + \Delta t \mathbf{c}_i \cdot \left(\frac{\partial h_i^n}{\partial \mathbf{x}}\right)_{FD} = \left(1 - \frac{\omega}{2}\right) h_i^n + \frac{\omega}{2} h_i^{eq,n}, \qquad (10)$$

where the subscript $n$ represents the $n^{th}$ time layer and the dimensionless relaxation frequency $\omega = \Delta t/\tau$. $h_i$ is the distribution function in terms of the discrete velocity $\mathbf{c}_i$, and its equilibrium is

$$h_i^{eq} = \frac{\delta\rho}{\rho_0} + \frac{\mathbf{c}_i \cdot \mathbf{u}}{RT_0}. \tag{11}$$

In Eq. (10), to remove numerical implicitness a new function is introduced [39],

$$H_i = h_i + \frac{\omega}{2}\left(h_i - h_i^{eq}\right). \tag{12}$$

Based on it, the fluid perturbations are now computed by

$$\delta\rho = \rho_0 \sum_i w_i H_i, \qquad \mathbf{u} = \sum_i w_i H_i \mathbf{c}_i. \tag{13}$$

As to the finite-difference approximations of the spatial gradients, $\left(\partial h_i^n / \partial \mathbf{x}\right)_{FD}$, the second-order upwind scheme (SUS) is used for those on the bulk nodes while those on the nodes next to solid boundaries are specified by the hybrid scheme (HS) consisting of the central difference and first-order upwind approaches [24]. For demonstration, I take the component of one gradient in the $y$ direction as an example. Its SUS approximation at a bulk node $(x_b, y_b)$ is

$$\left.\frac{\partial h_i^n}{\partial y}\right|_{x_b, y_b} \approx \frac{3h_i^n(x_b, y_b) - 4h_i^n(x_b, y_b - \Delta y) + h_i^n(x_b, y_b - 2\Delta y)}{2\Delta y}, \qquad c_{iy} > 0, \tag{14a}$$

$$\left.\frac{\partial h_i^n}{\partial y}\right|_{x_b, y_b} \approx -\frac{3h_i^n(x_b, y_b) - 4h_i^n(x_b, y_b + \Delta y) + h_i^n(x_b, y_b + 2\Delta y)}{2\Delta y}, \qquad c_{iy} < 0. \tag{14b}$$

When the gradient is at a node $(x_0, y_0)$ next to a solid boundary, it is approximated by

$$\left.\frac{\partial h_i^n}{\partial y}\right|_{x_0,y_0} \approx \frac{(1-\varepsilon)h_i^n(x_0,y_0+\Delta y)+2\varepsilon h_i^n(x_0,y_0)-(1+\varepsilon)h_i^n(x_0,y_0-\Delta y)}{2\Delta y}, \quad c_{iy}>0, \quad (14c)$$

$$\left.\frac{\partial h_i^n}{\partial y}\right|_{x_0,y_0} \approx \frac{(1+\varepsilon)h_i^n(x_0,y_0+\Delta y)-2\varepsilon h_i^n(x_0,y_0)-(1-\varepsilon)h_i^n(x_0,y_0-\Delta y)}{2\Delta y}, \quad c_{iy}<0. \quad (14d)$$

$\Delta y$ is the lattice spacing in the $y$ direction, and $\varepsilon$ is a numerical prefactor to tune the contribution from the first-order upwind and central difference approaches. To obtain stable and accurate results, this parameter is specified between 0.05 and 0.1. Noted that the velocity space in Eq. (10) has been tacitly discretized, whose specifications based on the HGH quadrature will be elaborated in the next section.

**B. HGH-based discrete velocity space**

As discussed in Introduction, the discrete velocities in a LB model are specified based on its moment evaluation using a designated quadrature scheme [7–10]. In the conventional LB theory, the distribution function is approximated as a polynomial series in the velocity space (either by the Chapman-Enskog or Grad expansion), and thus the relevant moments are reduced to integrals of polynomials in $\mathbf{c}$ with respect to the weight function $\exp(-\mathbf{c}^2/2RT_0)$. In these scenarios, the GH quadrature is the best choice for moment evaluation, and gives rise to a series of the GH-based discrete velocity spaces in the LB literature [7–10]. In this section, I take the moment evaluation in the 1D velocity space as an example to demonstrate such HGH-based velocity discretization. Under the isothermal conditions, the general $n^{th}$-order moment related to the linearized Boltzmann BGK equation is

$$M^{(n)} = \frac{1}{\sqrt{2\pi RT_0}} \int_{-\infty}^{+\infty} \exp\left(-\frac{c^2}{2RT_0}\right) h(c) c^n \, dc. \quad (15)$$

This moment can be split into two half-space integrals, each of which is evaluated by the HGH quadature,

$$M^{(n)} = \frac{1}{\sqrt{2\pi RT}} \int_0^{+\infty} \exp\left(-\frac{c^2}{2RT_0}\right) \left[h(-c)(-c)^n + h(c)c^n\right] dc$$
$$\approx \sum_{i=1}^{m} w_i h_{\bar{i}} (-c_i)^n + \sum_i w_i h_i c_i^n, \tag{16}$$

where $c_i$ and $w_i$ represent the discrete velocity and corresponding weight derived from the HGH quadrature. $h_{\bar{i}}$ is the distribution function in terms of $(-c_i)$. For 2D and 3D flows, the results of Eq. (16) can easily produce the corresponding discrete velocity spaces using tensor product [10]. Table I presents a series of 2D discrete velocity spaces using the HGH quadrature schemes with 3 to 10 nodes [29]. They, combined with Eqs. (10)–(14d), will form a hierachy of the linearized HGH-based LB models. For convenience, I denote these models by $D_2Q_N^H$ [40] with $N = 4m^2$, $m = 3,......10$ hereafter.

Table I. The 2D discrete velocity spaces based on the HGH quadrature with 3 to 10 nodes ($m = 3,......10$). $\xi_j^H$ and $w_j^H$ are the nodes and weights of the HGH quadrature. $\mathbf{c}_i$ and $W_i$ are the discrete velocities and moment weights in the resulting HGH-based LB models, $D_2Q_N^H$ with $N = 4m^2$. The subscript, *FS*, represents full symmetry.

| | |
|---|---|
| discrete velocities | $\mathbf{c}_i = \begin{cases} (\pm c_j, \pm c_j), & j = 1,2,......,m;\ i = 1,2,......,4m, \\ (c_j, c_k)_{FS}, & j,k = 1,2,......,m,\ j \neq k;\ i = 4m+1,......,N. \end{cases}$ <br><br> with $c_j = \sqrt{2RT_0}\,\xi_j^H$. |
| moment weights | $W_i = \begin{cases} w_j^2, & j = 1,2,......,m;\ i = 1,2,......,4m, \\ w_j \times w_k, & j,k = 1,2,......,m,\ j \neq k;\ i = 4m+1,......,N. \end{cases}$ <br><br> with $w_j = \dfrac{w_j^H}{2}$. |

## C. HGH-based lattice Boltzmann simulations of Couette flows

The linearized HGH-based LB models developed in Secs III A and B are now applied to simulate Couette flows at $Kn=10$ and $Kn=100$. To obtain the dimensionless results, I set the plate distance $L=1$, reference density $\rho_0=1$ and wall speed $u_w=1$. The Mach number in simulation $M=0.16$ and Courant-Friedrichs-Lewey number $CFL=0.2$. The discrete velocity versions of Eqs. (6) and (7) were employed as the boundary conditions on the plate walls while periodic boundary conditions were applied to specify $h$ at the two ends in the $x$ direction. To gain reasonable computational efficiency, all simulations were carried out on $N_x \times N_y = 4 \times 120$ grids. The numerical results are nearly $2^{nd}$-order accurate and grid independent.

Figure 2 shows the dimensionless streamwise velocities $U$ of Couette flow at $Kn=10$ obtained by the eight HGH-based LB models, together with the high-accuracy solution to the linearized Boltzmann BGK equation [31] solved by the method in Ref. [32]. Due to symmetry, only the results in the top half ($0.5 \leq Y \leq 1$) are exhibited. It is plain that the streamwise velocities obtained by $D_2Q_{36}^H$ and $D_2Q_{64}^H$ are in low accuracy–large errors are found in the Knudsen layer near the wall in these two cases. When $D_2Q_{100}^H$ and $D_2Q_{144}^H$ are used, such errors have been reduced and their results are closer to the high-accuracy solution to the linearized Boltzmann BGK equation. Unfortunately, further increasing AP does not significantly improve the accuracy of the LB results. Take $D_2Q_{196}^H$ and $D_2Q_{400}^H$ in Fig. 2(b) as examples. Although the latter has more than double discrete velocities of the former, their results almost collapse into a single line. Both deviate evidently from the high-accuracy solution to the linearized Boltzmann BGK equation.

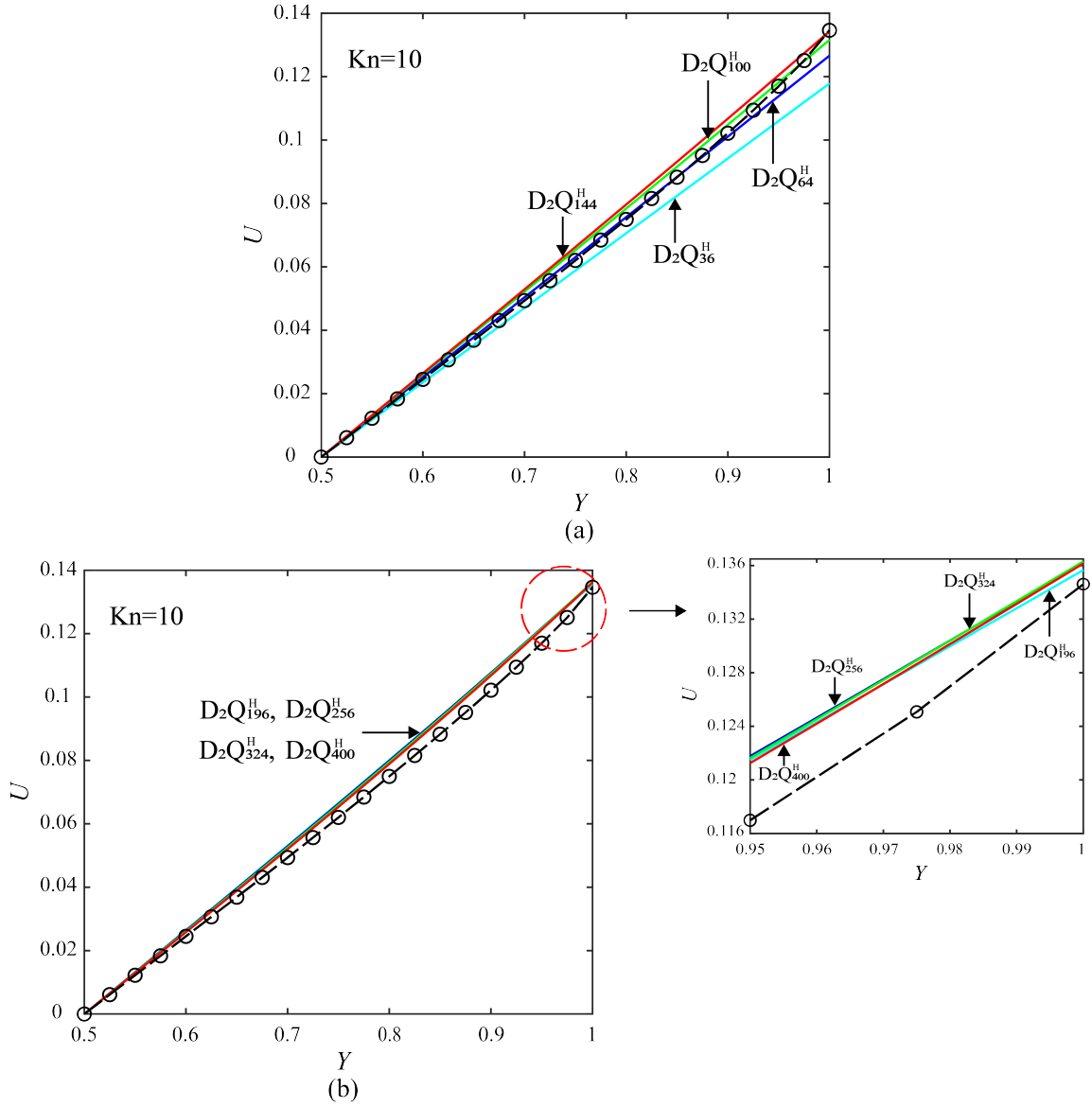

FIG. 2. Comparison of the streamwise velocities of Couette flow at $Kn=10$ by the HGH-based LB models (solid lines) with the high-accuracy solution to the linearized Boltzmann BGK equation (open cycles with dashed lines). (a): $D_2Q_{36}^H$, $D_2Q_{64}^H$, $D_2Q_{100}^H$ and $D_2Q_{144}^H$; (b): $D_2Q_{196}^H$, $D_2Q_{256}^H$, $D_2Q_{324}^H$ and $D_2Q_{400}^H$. Insert: The velocity details near the top wall by $D_2Q_{196}^H$, $D_2Q_{256}^H$, $D_2Q_{324}^H$ and $D_2Q_{400}^H$.

Figure 3 further shows the streamwise velocities of Couette flow at $Kn=100$. Generally, the results corresponding to a high AP HGH quadrature scheme are more accurate than their low AP counterparts. However, no HGH-based LB models in Fig. 3 are able to capture the strongly noncontinuum effects in this free molecular flow problem. In addition, I find that all velocity profiles in Figs. 2 and 3 are concave upward with the increasing slopes of their tangent lines along the $y$ direction. Nevertheless, the slopes by the LB results vary much more slowly than that of the high-accuracy solution, in particular in the Knudsen layer near the plate wall. These numerical simulations clearly demonstrate that the HGH-based LB models are ineffective to describe Couette flows at large Knudsen numbers; Their numerical errors cannot be removed even using a HGH quadrature scheme at high AP.

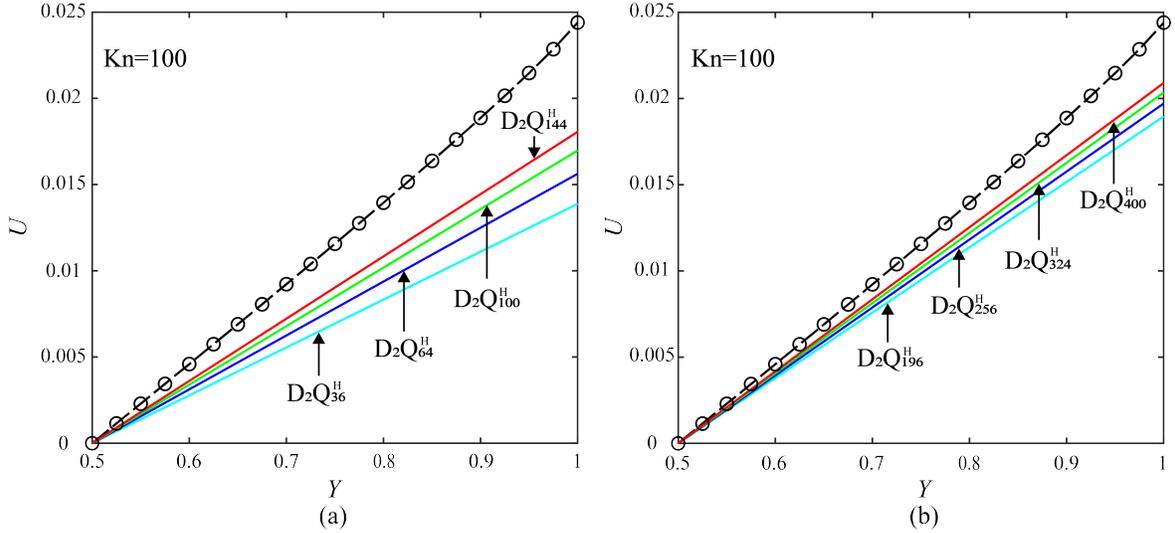

FIG. 3. Comparison of the streamwise velocities of Couette flow at $Kn=100$. Details as in Fig. 2

## IV. DISCUSSION ON THE EVALUATIONS OF THE ABRAMOWITZ FUNCTIONS

To figure out the reason for the poor accuracy of the HGH-based LB models, I revisit Section II B, where the streamwise velocity of Couette flow, $U$, is given by Eq. (9). It is clear that when the Knudsen number is large, this moment actually depends on two Abramowitz functions, $\mathscr{I}_0$ and $\mathscr{I}_{-1}$; It cannot be simply regarded as an integral of polynomials with the weight function $\exp(-\xi_y^2)$ in the velocity space. It is implied that the accuracy of a quadrature scheme for computing the streamwise velocity $U$ is determined by whether it can evaluate $\mathscr{I}_0$ and $\mathscr{I}_{-1}$ properly.

## A. Distributions of $I_0$, $I_{-1}$ and the HGH-based discrete velocities in the range of $\xi_y > 0$

Bearing this insight in mind, I first examine the accuracy of the HGH quadrature. To this end, the integrands of the two Abramowitz functions, $I_0$ and $I_{-1}$, are illustrated in the range of $\xi_y > 0$ in Figs 4 and 5. The distributions of the discrete velocities derived from the 4–, 7– and 10–node HGH quadrature schemes are also displayed for comparison. Note that for a succinct demonstration, Figs. 4 and 5 only exhibit the results at $Y = 0.1$ when $Kn = 10$ and 100, respectively.

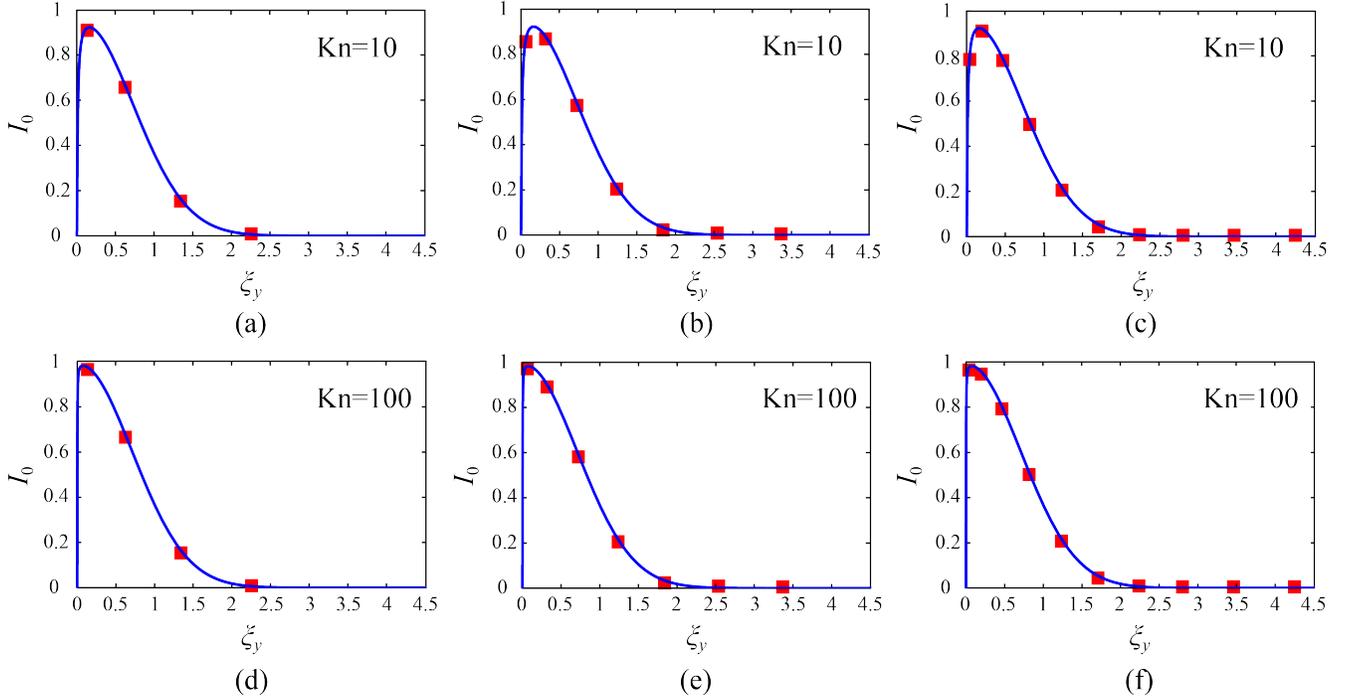

FIG. 4. The profiles of $I_0$ (blue lines) in the range of $\xi_y > 0$ at $Y = 0.1$ when $Kn = 10$ and 100. HGH-based discrete velocities (red squares) are also marked. (a) and (d): 4–node scheme $\left(D_2 Q_{64}^H\right)$; (b) and (e): 7–node scheme $\left(D_2 Q_{196}^H\right)$; (c) and (f): 10–node scheme $\left(D_2 Q_{400}^H\right)$.

It is clear that the profiles of $I_0$ and $I_{-1}$ are far from a Gaussian distribution in the velocity space in these two large-Kn cases. They have sharp peaks in a narrow vicinity of $\xi_y = 0$ and fast decaying tails when $\xi_y$ goes to the positive infinity. These profiles are asymmetric about any point in the range of $\xi_y > 0$, and have almost decayed to zero when $\xi_y \geq 4$. Importantly, the peak near $\xi_y = 0$ become more pronounced with the increasing Knudsen number, in particular in the case of $I_{-1}$. As to the HGH-based discrete velocities, only four samples are specified by the 4-node scheme, which are rather insufficient to characterize the variations of $I_0$ and $I_{-1}$ in the range of $\xi_y > 0$. It is thus not surprising that $D_2Q_{64}^H$ suffers from large errors in its simulations of Couette flows at $Kn = 10$ and 100. This situation has been ameliorated when the 7–node scheme was used. In comparison to the 4-node scheme, it introduces more discrete velocities, and many of them are in the range of $\xi_y$ where $I_0$ and $I_{-1}$ have large variations. As a consequence, the accuracy of $D_2Q_{196}^H$ has been substantially improved. Unfortunately, such an improvement is not enhanced when the number of discrete velocities increases up to 10. See Figs. 4(c), 4(f), 5(c) and 5(f)–the majority of discrete velocities specified by the 10-node scheme are allocated in the tails of $I_0$ and $I_{-1}$, playing a trivial role in evaluating the values of $\mathscr{I}_0$ and $\mathscr{I}_{-1}$. The results in Figs. 4 and 5 reveal that the discrete velocities derived from the HGH quadrature fail to capture the variation features of $I_0$ and $I_{-1}$ in $\xi_y > 0$. In the next session, a different quadrature scheme is employed to attack this issue, and the distribution of the resulting discrete velocities are investigated for Couette flows at large Knudsen numbers.

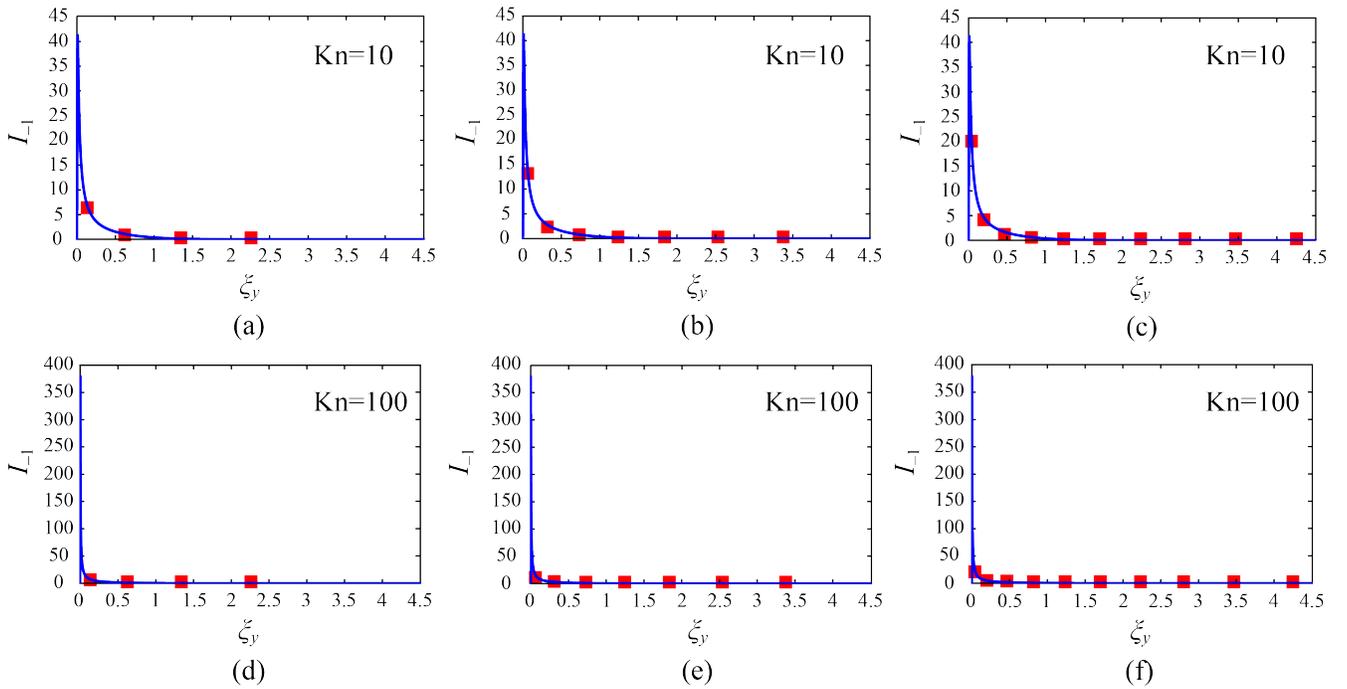

FIG. 5. The profiles of $I_{-1}$ in the range of $\xi_y > 0$. Details as in Fig. 4.

## B. Distributions of $I_0$, $I_{-1}$ and the GL-based discrete velocities in the range of $\xi_y > 0$

Figures. 4 and 5 illustrate that the integrands, $I_0$ and $I_{-1}$, decay rapidly at large $\xi_y$. This implies that their tails beyond a certain $\xi_y$ can be safely truncated while without significant errors in the evaluations of $\mathcal{I}_0$ and $\mathcal{I}_{-1}$. It follows that

$$\mathcal{I}_0\left(\frac{aY}{Kn}\right) \approx \int_0^{\xi_0} I_0\left(\xi_y, \frac{aY}{Kn}\right) d\xi_y, \qquad \mathcal{I}_{-1}\left(\frac{aY}{Kn}\right) \approx \int_0^{\xi_0} I_{-1}\left(\xi_y, \frac{aY}{Kn}\right) d\xi_y, \qquad (17)$$

Here the upper limit $\xi_0 = 4$. Next, the Möbius transformation is applied,

$$\xi^L = \frac{3\xi_y - 2}{2(\xi_y + 1)}, \qquad (18)$$

resulting in Eq. (17) to be

$$\mathcal{I}_0\left(\frac{aY}{Kn}\right) \approx \int_{-1}^{1} \frac{10}{(3-2\xi^L)^2} \exp\left[-\frac{4(1+\xi^L)^2}{(3-2\xi^L)^2} - \frac{aY}{2Kn}\frac{(3-2\xi^L)}{(1+\xi^L)}\right] d\xi^L, \qquad (19)$$

and

$$\mathcal{I}_{-1}\left(\frac{aY}{Kn}\right) \approx \int_{-1}^{1} \frac{5}{(1+\xi^L)(3-2\xi^L)} \exp\left[-\frac{4(1+\xi^L)^2}{(3-2\xi^L)^2} - \frac{aY}{2Kn}\frac{(3-2\xi^L)}{(1+\xi^L)}\right] d\xi^L. \qquad (20)$$

Note that Eqs. (19) and (20) are now two integrals in a finite range of $\xi_y$ between $-1$ and $1$. They can thus be well evaluated using the GL quadrature [26],

$$\mathcal{I}_0\left(\frac{aY}{Kn}\right) \approx \sum_i w_i \exp\left(-\frac{aY}{Kn}\frac{1}{\xi_{y_i}}\right), \qquad \mathcal{I}_{-1}\left(\frac{aY}{Kn}\right) \approx \sum_i \frac{w_i}{\xi_{y_i}} \exp\left(-\frac{aY}{Kn}\frac{1}{\xi_{y_i}}\right), \qquad (21)$$

where $\xi_{y_i} = 2(1+\xi_i^L)/(3-2\xi_i^L)$ and $w_i = 2w_i^L(\xi_i + 1)^2 \exp(-\xi_i^2)/5$. $\xi_i^L$ and $w_i^L$ are the nodes and weights of the GL quadrature [26].

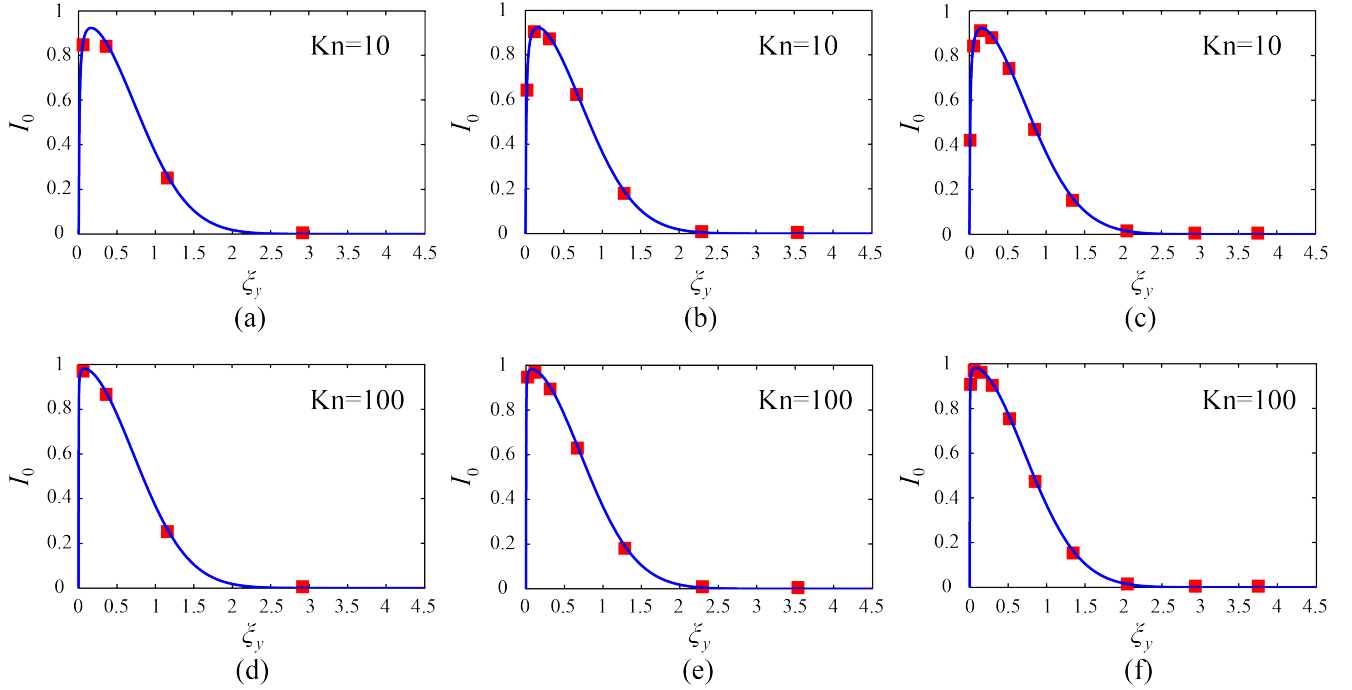

FIG. 6. The profiles of $I_0$ (blue lines) in the range of $\xi_y > 0$ at $Y = 0.1$ when $Kn = 10$ and $100$. GL-based discrete velocities (red squares) are also marked. (a) and (d): 4−node scheme; (b) and (e): 7−node scheme; (c) and (f): 10−node scheme.

I apply Eq. (21) to generate the discrete velocities, and exhibit their distributions in the range of $\xi_y > 0$. Figures 6 and 7 show the results from the 4−, 7− and 10−node GL quadrature schemes, together with the profiles of $I_0$ and $I_{-1}$ at $Y = 0.1$ when $Kn = 10$ and $100$. In comparison with the HGH-based results in Figs. 4 and 5, the discrete velocities generated by the GL quadrature schemes are much closer to the singularity point $\xi_y = 0$; Many of them fall into the range of $\xi_y$, where $I_0$ and $I_{-1}$ have dramatic variations. Apparently, the distributions of these GL-based discrete velocities are much more consistent with the variations of $I_0$ and $I_{-1}$ at large Knudsen numbers. Inspired by these encouraging results, here I prompt a quick question: Is the LB method equipped with these GL-based discrete velocities in favor of simulating strongly noncontinuum Couette flows in high accuracy?

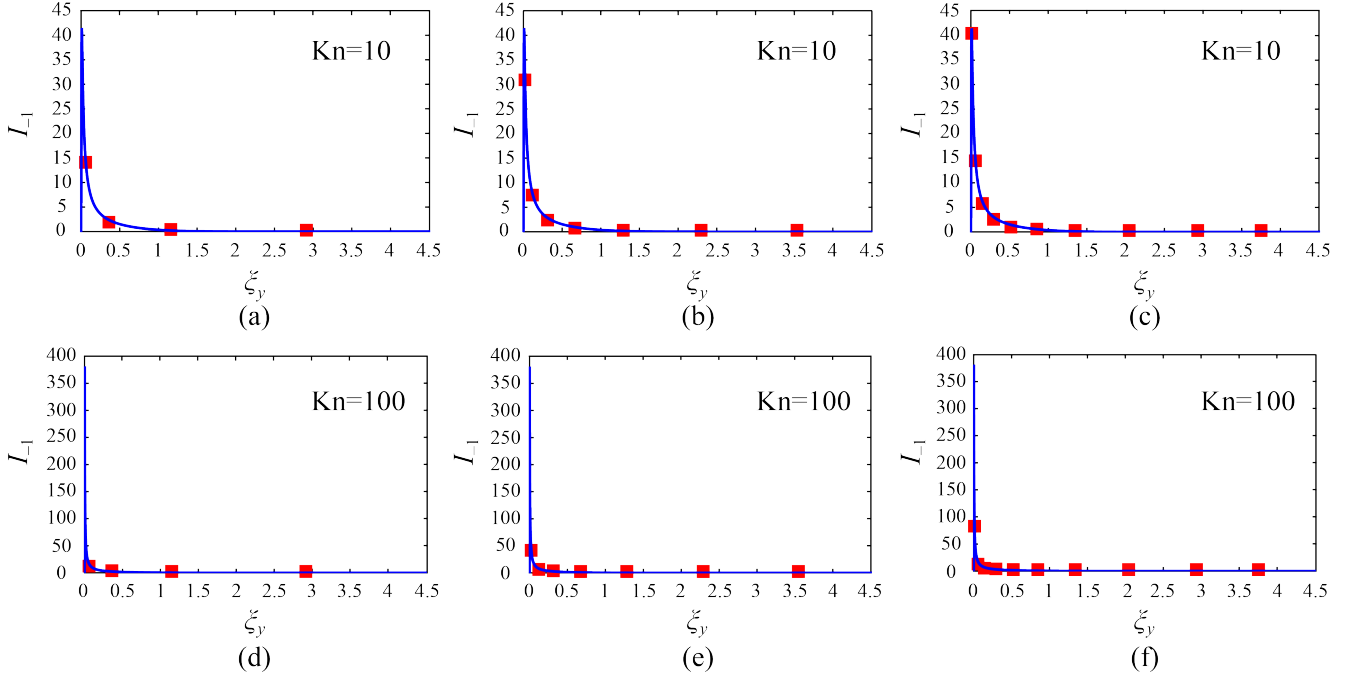

FIG. 7. The profiles of $I_{-1}$ in the range of $\xi_y > 0$. Details as in Fig. 6.

## V. GUASS-LEGENDRE BASED LINEARIZED LATTICE BOLTZMANN MODEL FOR NONCONTINUUM BOUNDED GAS FLOWS

**A. GL-based discrete velocity space and lattice Boltzmann model**

To answer the question at the end of Sec. IV B, a hierarchy of the linearized LB models based on the GL-based discrete velocities is constructed. The derivations are similar to those in Section III B, which start from Eq. (15). Through use of the range truncation, Eq. (17), Möbius transformation, Eq. (18), and evaluations by the GL quadrature, the moment integral of Eq. (15) is approximated by

$$M^{(n)} = \frac{1}{\sqrt{2\pi RT}} \int_{-\infty}^{+\infty} \exp\left(-\frac{c^2}{2RT_0}\right) h(c) c^n \, dc \approx \sum_{i=1}^{m} w_i h_{\bar{i}} \left(-c_i\right)^n + \sum_i w_i h_i c_i^n, \qquad (22)$$

where

$$c_i = 2\sqrt{2RT_0}\,\frac{1+\xi_i^L}{3-2\xi_i^L}, \qquad (23)$$

and

$$w_i = \frac{2w_i^L}{5\sqrt{\pi}} \exp\left(-\frac{c_i^2}{2RT_0}\right)\left(\frac{c_i}{\sqrt{2RT_0}}+1\right)^2. \qquad (24)$$

Again, the 2D discrete velocity spaces are produced using tensor product based on Eqs. (23) and (24). Table II presents their specifications corresponding to the GL quadrature schemes with 3 to 10 nodes. To be similar with Section III B, these discrete velocity spaces, together with Eqs. (10), (11) and (14), form the GL-based LB models, $D_2Q_N^L$ [40], with $N = 4m^2$, $m = 3,......10$.

Table II. The 2D discrete velocity spaces based on the GL quadrature with 3 to 10 nodes ($m = 3,......,10$). $\xi_j^L$ and $w_j^L$ are the nodes and weights of the GL quadrature. $\mathbf{c}_i$ and $W_i$ are the discrete velocities and moment weights in the resulting GL-based LB models, $D_2Q_N^L$ with $N = 4m^2$. The subscript, *FS*, represents full symmetry.

| | |
|---|---|
| discrete velocities | $\mathbf{c}_i = \begin{cases} (\pm c_j, \pm c_j), & j=1,2,......,m;\ i=1,2,......,4m, \\ (c_j, c_k)_{FS}, & j,k=1,2,......m,\ j\neq k;\ i=4m+1,......,N. \end{cases}$ <br><br> with $c_j = 2\sqrt{2RT_0}\left(1+\xi_j^L\right)/\left(3-2\xi_j^L\right)$. |
| moment weights | $W_i = \begin{cases} w_j^2, & j=1,2,......,m;\ i=1,2,......,4m, \\ w_j \times w_k, & j,k=1,2,......m,\ j\neq k;\ i=4m+1,......,N. \end{cases}$ <br><br> with $w_j = \frac{2w_j^L}{5\sqrt{\pi}} \exp\left(-\frac{c_j^2}{2RT_0}\right)\left(\frac{c_j}{\sqrt{2RT_0}}+1\right)^2$. |

## B. GL-based lattice Boltzmann simulations of Couette flows

The hierarchy of the GL-based LB models obtained in Section V A was applied to simulate Couette flows at $Kn = 10$ and $Kn = 100$. The numerical settings were specified the same as those in Section III C, except that the discrete particle velocities were replaced by those in Table II.

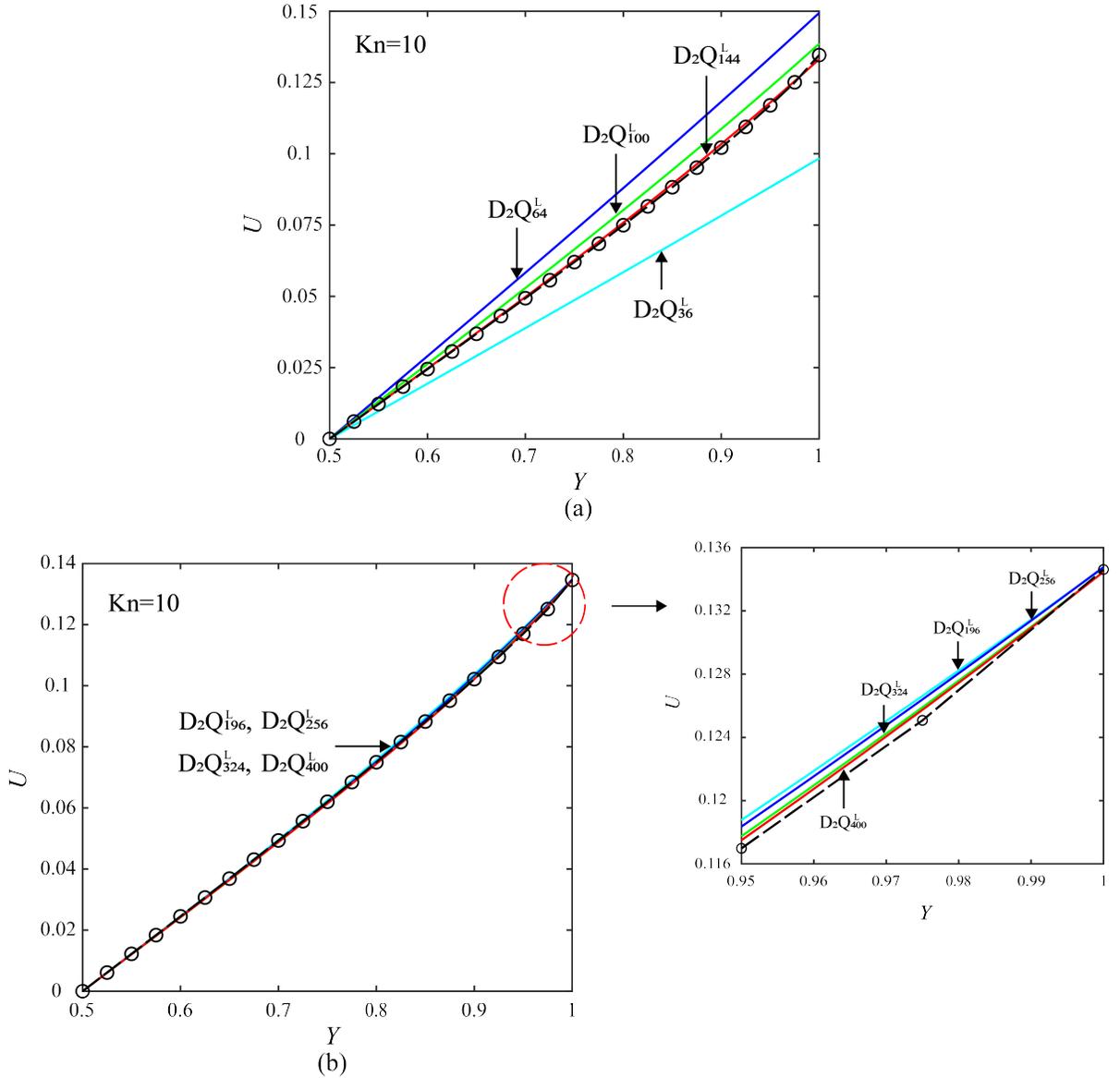

FIG. 8. Comparison of the streamwise velocities of Couette flow at $Kn=10$ by the GL-based LB models (solid lines) with the high-accuracy solution to the linearized Boltzmann BGK equation (open cycles with dashed lines). (a): $D_2Q_{36}^L$, $D_2Q_{64}^L$, $D_2Q_{100}^L$ and $D_2Q_{144}^L$; (b): $D_2Q_{196}^L$, $D_2Q_{256}^L$, $D_2Q_{324}^L$ and $D_2Q_{400}^L$. Insert: The velocity details near the top wall by $D_2Q_{196}^L$, $D_2Q_{256}^L$, $D_2Q_{324}^L$ and $D_2Q_{400}^L$.

Figure 8 shows the streamwise velocities of Couette flow at $Kn=10$ obtained by the GL-based LB models. It is striking that the LB models based on the GL schemes at high AP (from $D_2Q_{196}^L$ to $D_2Q_{400}^L$) obtained rather accurate results in this strong transition flow. In comparison to the high-accuracy solution to the Boltzmann BGK equation, the overall root-mean-squared error of the results by $D_2Q_{196}^L$ is just $8.965\times10^{-4}$. It has further decreased to $4.016\times10^{-4}$ when $D_2Q_{400}^L$ is used. In particular, for the streamwise velocity in the Knudsen layer, its upward concavity and the slop changes of its tangent lines are all well captured by $D_2Q_{196}^L$, $D_2Q_{256}^L$, $D_2Q_{324}^L$ and $D_2Q_{400}^L$. To my best knowledge, the results in such high accuracy have never been reported in the previous HGH-based LB simulations. In Fig. 8, the insert zooms in the velocity details near the top wall by $D_2Q_{196}^L$, $D_2Q_{256}^L$, $D_2Q_{324}^L$ and $D_2Q_{400}^L$. It is found that the accuracy of the LB results improves gradually with the increasing AP. This is in accord with the findings in Figs. 6 and 7−when AP of the GL quadrature grows, more discrete velocities will be generated and allocated in the neighborhood of $\xi_y = 0$, leading to better evaluations of $\mathscr{I}_0$ and $\mathscr{I}_{-1}$.

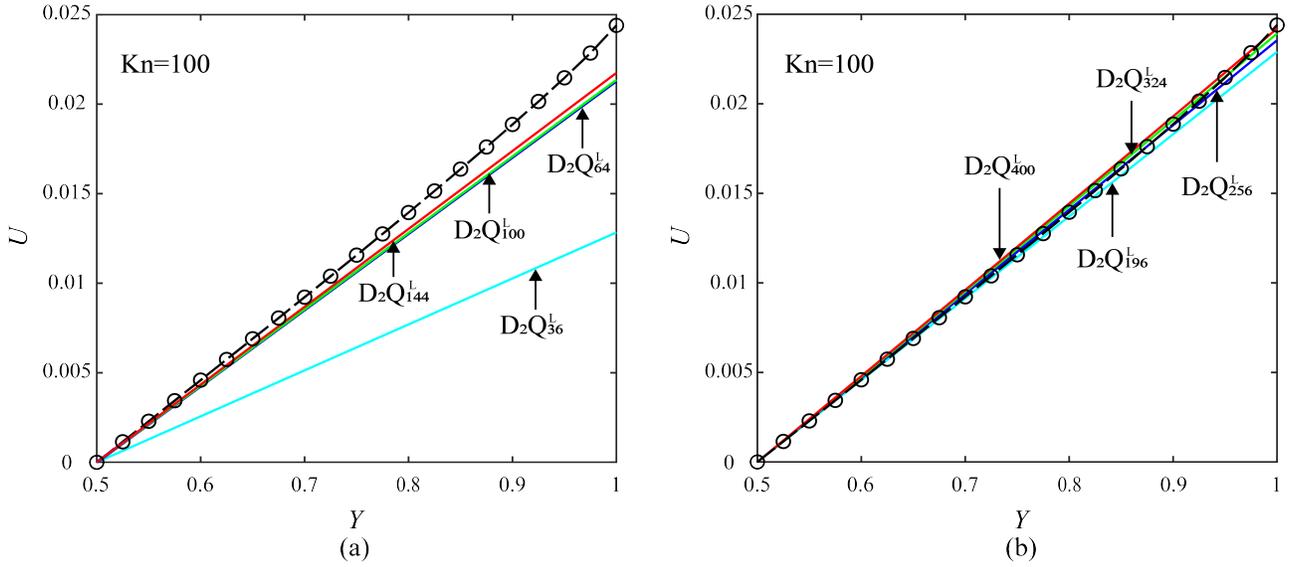

FIG. 9. Comparison of the streamwise velocities of Couette flow at $Kn = 100$. Details as in Fig. 8.

Simulations by these GL-based LB models were also conducted for Couette flow at $Kn = 100$, and their numerical results are shown in Fig. 9. In this free molecular flow, the streamwise velocities by the models corresponding to low AP suffer from large errors. The worst is the results by $D_2Q_{36}^L$, and those by $D_2Q_{64}^L$, $D_2Q_{100}^L$ and $D_2Q_{144}^L$ also deviate significantly from the high-accuracy solution to the Boltzmann BGK equation. Accuracy is improved in the simulations of $D_2Q_{196}^L$, $D_2Q_{256}^L$, $D_2Q_{324}^L$ and $D_2Q_{400}^L$. As shown in Fig. 9(b), the streamwise velocity by $D_2Q_{400}^L$ is well agreed with the high-accuracy solution to the Boltzmann BGK equation, although they don't exactly overlap each other. Note that in comparison with the results at $Kn = 10$ in Fig. 8, the accuracy of the GL-based LB models has slightly decreased in this free molecular flow. This is understandable since $I_0$ and $I_{-1}$ have extremely sharp peaks in a rather small vicinity of $\xi_y = 0$, when the Knudsen number jumps to 100. Under this circumstance, only a very few GL-based discrete velocities can be included in such a narrow $\xi_y$-range.

## VI. CONCLUSION

So far, both the HGH- and GL-based LB models have been applied to simulate noncontinuum Couette flows at $Kn = 10$ and $Kn = 100$. Unfortunately, large errors are found in all HGH-based LB simulations, although AP of some HGH quadrature schemes has increased up to a very large value. From the theoretical point of view, these failures dwell in the premise underlying the current LB theory, which assumes the distribution function continuous in the velocity space, and represents it by the local Maxwellian times a polynomial expansion. Each term in the latter is further specified as a product with separate dependence on $(t, \mathbf{x})$ and $\mathbf{c}$ [10,28]. Nevertheless, discontinuity of the distribution function in $\mathbf{c}$ does exist at the flat walls in Couette flow. This singularity leads to the behaviors of distribution function in the Knudsen layer deviating from the aforementioned polynomial approximations [27,30]. It is therefore not suprising that in the strongly noncontinuum flows where the Knudsen layer plays a dominant role, the HGH quadarature fails to give exact or even high-accuracy evaluations of the relevant moments.

As an alternative, this article focuses on the streamwise velocity of Couette flow, which was directly solved from the linearized Boltzmann BGK equation without making use of a *priori* polynomial expansion of the distribution function. The GL quadrature was proposed to evaluate the involved Abramowitz functions, and led to essentially distinct velocity discretization. The resulting LB models, especially those at high AP, have achieved unexpected but rather accurate results of Couette flows in the strong transition and even free molecular flow regimes. These numerical simulations demonstrate the proposed velocity discretization and LB models can well capture strongly noncontinuum flow features caused by flat solid boundaries. Future studies will continue refining these models, and investigate their applications to noncontinuum gas flows bounded in more complex micro and nanogeometries.

## ACKNOWLEDGEMENTS

The author would like to acknowledge support from Zhejiang Provincial Natural Science Foundation of China (Grant No. LY16E060001).

# APPENDIX NODES AND WEIGHTS OF THE HGH QUADRATURE AND GL QUADRATURE

In this appendix, the nodes and weights of the HGH quadrature and GL quadrature are given in two tables for readers' reference.

Table A. The nodes and weights of the HGH quadrature for evaluating

$$\int_0^{+\infty} \frac{2}{\sqrt{\pi}} \exp(-\xi^2) f(\xi) \, d\xi \approx \sum_j w_j^H f(\xi_j^H) \ [29].$$

|  | Nodes, $\xi_j^H$ | Weights, $w_j^H$ |
|---|---|---|
| $i = 3$ | 1.90554149798192×10$^{-1}$<br>8.48251867544577×10$^{-1}$<br>1.79977657841573 | 5.03290700898970×10$^{-1}$<br>4.47366532895383×10$^{-1}$<br>4.93427662056466×10$^{-2}$ |
| $i = 4$ | 1.33776446996068×10$^{-1}$<br>6.24324690187190×10$^{-1}$<br>1.34253782564499<br>2.26266447701036 | 3.67065127919384×10$^{-1}$<br>4.75168480845835×10$^{-1}$<br>1.50573737408561×10$^{-1}$<br>7.19265382622035×10$^{-3}$ |
| $i = 5$ | 1.00242151968216×10$^{-1}$<br>4.82813966046201×10$^{-1}$<br>1.06094982152572<br>1.77972941852026<br>2.66976035608766 | 2.80296326927255×10$^{-1}$<br>4.42698202214928×10$^{-1}$<br>2.38559884611999×10$^{-1}$<br>3.75148389159569×10$^{-2}$<br>9.30747329860493×10$^{-4}$ |
| $i = 6$ | 7.86006594130979×10$^{-2}$<br>3.86739410270631×10$^{-1}$<br>8.66429471682044×10$^{-1}$<br>1.46569804966352<br>2.17270779693900<br>3.03682016932287 | 2.22121072870846×10$^{-1}$<br>3.93978327105124×10$^{-1}$<br>2.90286283564440×10$^{-1}$<br>8.57716408753570×10$^{-2}$<br>7.73156223123715×10$^{-3}$<br>1.11113352995475×10$^{-4}$ |
| $i = 7$ | 6.37164846067008×10$^{-2}$<br>3.18192018888619×10$^{-1}$<br>7.24198989258373×10$^{-1}$<br>1.23803559921509<br>1.83852822027095<br>2.53148815132768<br>3.37345643012458 | 1.81228938702362×10$^{-1}$<br>3.45644889994293×10$^{-1}$<br>3.10899086759459×10$^{-1}$<br>1.36116596851485×10$^{-1}$<br>2.47027998304411×10$^{-2}$<br>1.39518072944905×10$^{-3}$<br>1.25071325106075×10$^{-5}$ |
| $i = 8$ | 5.29786439318511×10⁻² <br> 2.67398372167765×10⁻¹ <br> 6.16302884182400×10⁻¹ <br> 1.06424631211622 <br> 1.58885586227006 <br> 2.18392115309586 <br> 2.86313388370807 <br> 3.68600716272440 | 1.51326014366857×10⁻¹ <br> 3.02778833237947×10⁻¹ <br> 3.11380065379352×10⁻¹ <br> 1.77661362002010×10⁻¹ <br> 5.05673080227936×10⁻² <br> 6.05706687740607×10⁻³ <br> 2.28004412108310×10⁻⁴ <br> 1.34570152671232×10⁻⁶ |
| $i = 9$ | 4.49390308011905×10⁻² <br> 2.28605305560523×10⁻¹ <br> 5.32195844331623×10⁻¹ <br> 9.27280745338049×10⁻¹ <br> 1.39292385519585 <br> 1.91884309919740 <br> 2.50624783400570 <br> 3.17269213348120 <br> 3.97889886978974 | 1.28735617216578×10⁻¹ <br> 2.66230673484828×10⁻¹ <br> 3.00628954027932×10⁻¹ <br> 2.06777377033598×10⁻¹ <br> 8.05031389339135×10⁻² <br> 1.57763412665662×10⁻² <br> 1.31326716370170×10⁻³ <br> 3.44911902492407×10⁻⁵ <br> 1.39682634077374×10⁻⁷ |
| $i = 10$ | 3.87385243256994×10⁻² <br> 1.98233304012949×10⁻¹ <br> 4.65201111814507×10⁻¹ <br> 8.16861885591907×10⁻¹ <br> 1.23454132402774 <br> 1.70679814968865 <br> 2.22994008892444 <br> 2.80910374689825 <br> 3.46387241949537 <br> 4.25536180636561 | 1.11204133714046×10⁻¹ <br> 2.35467982990322×10⁻¹ <br> 2.84409874226013×10⁻¹ <br> 2.24191270127519×10⁻¹ <br> 1.09676675317101×10⁻¹ <br> 3.04937885088302×10⁻² <br> 4.29308737213101×10⁻³ <br> 2.58270468287074×10⁻⁴ <br> 4.90319654442849×10⁻⁶ <br> 1.40792060402115×10⁻⁸ |
| | Quadrature Nodes, $\xi_i^L$ | Quadrature Weights, $w_i^L$ |
|---|---|---|
| $i = 8$ | $5.29786439318511 \times 10^{-2}$ <br> $2.67398372167765 \times 10^{-1}$ <br> $6.16302884182400 \times 10^{-1}$ <br> $1.06424631211622$ <br> $1.58885586227006$ <br> $2.18392115309586$ <br> $2.86313388370807$ <br> $3.68600716272440$ | $1.51326014366857 \times 10^{-1}$ <br> $3.02778833237947 \times 10^{-1}$ <br> $3.11380065379352 \times 10^{-1}$ <br> $1.77661362002010 \times 10^{-1}$ <br> $5.05673080227936 \times 10^{-2}$ <br> $6.05706687740607 \times 10^{-3}$ <br> $2.28004412108310 \times 10^{-4}$ <br> $1.34570152671232 \times 10^{-6}$ |
| $i = 9$ | $4.49390308011905 \times 10^{-2}$ <br> $2.28605305560523 \times 10^{-1}$ <br> $5.32195844331623 \times 10^{-1}$ <br> $9.27280745338049 \times 10^{-1}$ <br> $1.39292385519585$ <br> $1.91884309919740$ <br> $2.50624783400570$ <br> $3.17269213348120$ <br> $3.97889886978974$ | $1.28735617216578 \times 10^{-1}$ <br> $2.66230673484828 \times 10^{-1}$ <br> $3.00628954027932 \times 10^{-1}$ <br> $2.06777377033598 \times 10^{-1}$ <br> $8.05031389339135 \times 10^{-2}$ <br> $1.57763412665662 \times 10^{-2}$ <br> $1.31326716370170 \times 10^{-3}$ <br> $3.44911902492407 \times 10^{-5}$ <br> $1.39682634077374 \times 10^{-7}$ |
| $i = 10$ | $3.87385243256994 \times 10^{-2}$ <br> $1.98233304012949 \times 10^{-1}$ <br> $4.65201111814507 \times 10^{-1}$ <br> $8.16861885591907 \times 10^{-1}$ <br> $1.23454132402774$ <br> $1.70679814968865$ <br> $2.22994008892444$ <br> $2.80910374689825$ <br> $3.46387241949537$ <br> $4.25536180636561$ | $1.11204133714046 \times 10^{-1}$ <br> $2.35467982990322 \times 10^{-1}$ <br> $2.84409874226013 \times 10^{-1}$ <br> $2.24191270127519 \times 10^{-1}$ <br> $1.09676675317101 \times 10^{-1}$ <br> $3.04937885088302 \times 10^{-2}$ <br> $4.29308737213101 \times 10^{-3}$ <br> $2.58270468287074 \times 10^{-4}$ <br> $4.90319654442849 \times 10^{-6}$ <br> $1.40792060402115 \times 10^{-8}$ |

Table B. The nodes and weights of the GL quadrature for evaluating $\int_{-1}^{1} f(\xi) \, d\xi \approx \sum_i w_i^L f(\xi_i^L)$ [26].

| | Quadrature Nodes, $\xi_i^L$ | Quadrature Weights, $w_i^L$ |
|---|---|---|
| $i = 3$ | $0$ <br> $\pm 7.74596669241483 \times 10^{-1}$ | $8.88888888888889 \times 10^{-1}$ <br> $5.55555555555556 \times 10^{-1}$ |
| $i = 4$ | $\pm 3.39981043584856 \times 10^{-1}$ <br> $\pm 8.61136311594053 \times 10^{-1}$ | $6.52145154862546 \times 10^{-1}$ <br> $3.47854845137454 \times 10^{-1}$ |
| $i = 5$ | $0$ <br> $\pm 5.38469310105683 \times 10^{-1}$ <br> $\pm 9.06179845938664 \times 10^{-1}$ | $5.68888888888889 \times 10^{-1}$ <br> $4.78628670499366 \times 10^{-1}$ <br> $2.36926885056189 \times 10^{-1}$ |
| $i = 6$ | $\pm 2.38619186083197 \times 10^{-1}$<br>$\pm 6.61209386466265 \times 10^{-1}$<br>$\pm 9.32469514203152 \times 10^{-1}$ | $4.67913934572691 \times 10^{-1}$<br>$3.60761573048139 \times 10^{-1}$<br>$1.71324492379170 \times 10^{-1}$ |
| $i = 7$ | $0$<br>$\pm 4.05845151377397 \times 10^{-1}$<br>$\pm 7.41531185599394 \times 10^{-1}$<br>$\pm 9.49107912342759 \times 10^{-1}$ | $4.17959183673469 \times 10^{-1}$<br>$3.81830050505119 \times 10^{-1}$<br>$2.79705391489277 \times 10^{-1}$<br>$1.29484966168870 \times 10^{-1}$ |
| $i = 8$ | $\pm 1.83434642495650 \times 10^{-1}$<br>$\pm 5.25532409916329 \times 10^{-1}$<br>$\pm 7.96666477413627 \times 10^{-1}$<br>$\pm 9.60289856497536 \times 10^{-1}$ | $3.62683783378362 \times 10^{-1}$<br>$3.13706645877887 \times 10^{-1}$<br>$2.22381034453374 \times 10^{-1}$<br>$1.01228536290376 \times 10^{-1}$ |
| $i = 9$ | $0$<br>$\pm 3.24253423403809 \times 10^{-1}$<br>$\pm 6.13371432700590 \times 10^{-1}$<br>$\pm 8.36031107326636 \times 10^{-1}$<br>$\pm 9.68160239507626 \times 10^{-1}$ | $3.30239355001260 \times 10^{-1}$<br>$3.12347077040003 \times 10^{-1}$<br>$2.60610696402935 \times 10^{-1}$<br>$1.80648160694857 \times 10^{-1}$<br>$8.12743883615744 \times 10^{-2}$ |
| $i = 10$ | $\pm 1.48874338981631 \times 10^{-1}$<br>$\pm 4.33395394129247 \times 10^{-1}$<br>$\pm 6.79409568299024 \times 10^{-1}$<br>$\pm 8.65063366688985 \times 10^{-1}$<br>$\pm 9.73906528517172 \times 10^{-1}$ | $2.95524224714753 \times 10^{-1}$<br>$2.69266719309996 \times 10^{-1}$<br>$2.19086362515982 \times 10^{-1}$<br>$1.49451349150581 \times 10^{-1}$<br>$6.66713443086881 \times 10^{-2}$ |

Through use of the rules specified in Table I and II, the quadrature nodes and weights in Table A and B will construct a series of discrete velocity spaces corresponding to different quadrature AP for the LB models.